\documentclass{article}
\usepackage{spconfa4,amsmath,graphicx, comment}
\usepackage{comment}
\usepackage{cite}
\usepackage{amsmath,amssymb,amsfonts}
\usepackage{algorithmic}
\usepackage{graphicx}
\usepackage{textcomp}
\usepackage{xcolor}
\def\BibTeX{{\rm B\kern-.05em{\sc i\kern-.025em b}\kern-.08em
		T\kern-.1667em\lower.7ex\hbox{E}\kern-.125emX}}

\definecolor{alorange}{RGB}{237,141,5}

\usepackage{balance}
\let\OLDthebibliography\thebibliography
\renewcommand\thebibliography[1]{
	\OLDthebibliography{#1}
	\setlength{\parskip}{0.2pt}
	\setlength{\itemsep}{0.1pt}
	\setlength{\baselineskip}{9.2pt}
}

\usepackage{booktabs}
\usepackage{hhline}
\usepackage{multirow}
\usepackage{makecell}
\usepackage{diagbox}

\usepackage{tikz}
\usetikzlibrary{shapes,arrows,calc,fit,automata,positioning}
\usepackage{amsmath,amsfonts,amssymb}
\usepackage{pdfpages}
\usetikzlibrary{intersections,backgrounds}
\usetikzlibrary{calc}
\usetikzlibrary{decorations.pathreplacing,calligraphy}

\usepackage{subcaption}
\usepackage{pgfplots}
\usepgfplotslibrary{polar}
\usepgfplotslibrary{fillbetween}
\pgfplotsset{compat=1.12}

\title{Neural Directional Filtering:\\Far-Field Directivity Control with a Small  Microphone Array}

\name{Julian Wechsler, Srikanth Raj Chetupalli, Mhd Modar Halimeh, Oliver Thiergart, Emanu{\"e}l A. P. Habets}
\address{International Audio Laboratories Erlangen\textsuperscript{$\ast$}, Am Wolfsmantel 33, 91058 Erlangen, Germany\thanks{\textsuperscript{$\ast$}A joint institution of Fraunhofer IIS and Friedrich-Alexander-Universit{\"a}t Erlangen-N{\"u}rnberg (FAU), Germany.}\vspace{-1em}\thanks{The authors gratefully acknowledge the scientific support and HPC resources provided by the National High Performance Computing Center (NHR@FAU) of the FAU under the NHR project b183dc. NHR funding is provided by federal and Bavarian state authorities. NHR@FAU hardware is partially funded by the German Research Foundation (DFG) – 440719683.}}
\usepackage[hidelinks]{hyperref} 

\usepackage[nolist]{acronym} 
\begin{acronym}[iSTFT]
	\acro{CDMA}[CDMA]{circular \ac{DMA}}
	
	\acro{sdri}[$\Delta$SDR]{improvement in \ac{SDR} over the unprocessed signal}
	\acro{DMA}[DMA]{differential microphone array}
	\acro{DNN}[DNN]{deep neural network}
	\acro{DOA}[DOA]{direction-of-arrival}
	\acrodefplural{DOA}[DOAs]{directions-of-arrival}
	
	\acro{iSTFT}[iSTFT]{inverse short-time Fourier transform}
	
	\acro{LDMA}[LDMA]{linear \ac{DMA}}
        \acro{LS}{least-squares}
	\acro{LSTM}[LSTM]{long short-term memory}
	
	\acro{RIR}[RIR]{room impulse response}
	\acro{DPIR}[DPIR]{direct-path impulse response}
 
	\acro{SDR}[SDR]{signal-to-distortion ratio}
        \acro{noisySDR}[reference microphone]{\ac{SDR} of the unprocessed omnidirectional reference microphone}
    \acrodefplural{noisySDR}[$\textrm{SDRs}^\textrm{omni}$]{\acp{SDR} of the unprocessed omnidirectional microphone}
	\acro{SNR}[SNR]{signal-to-noise ratio}
	\acro{STFT}[STFT]{short-time Fourier transform}
	
	\acro{TF}[TF]{time-frequency}
	\acro{tsdr}[SA-$\varepsilon$-tSDR]{source-aggregated and regularized thresholded \ac{SDR}}
	
	\acro{UCA}[UCA]{uniform circular array}

    \acro{VDM}[VDM]{virtual directional microphone}
\end{acronym}

\begin{document}
	\ninept
	\maketitle
	\begin{abstract}
		Capturing audio signals with specific directivity patterns is essential in speech communication.
		This study presents a \ac{DNN}-based approach to directional filtering,  alleviating the need for explicit signal models.
		More specifically, our proposed method uses a \ac{DNN} to estimate a single-channel complex mask from the signals of a microphone array. This mask is then applied to a reference microphone to render a signal that exhibits a desired directivity pattern.
		We investigate the training dataset composition and its effect on the directivity realized by the \ac{DNN} during inference.
		Using a relatively small \ac{DNN}, the proposed method is found to approximate the desired directivity pattern closely.
		Additionally, it allows for the realization of higher-order directivity patterns using a small number of microphones, which is a difficult task for linear and parametric directional filtering.
	\end{abstract}
	\vspace{-0.5em}
    \section{Introduction}
	\label{sec:intro}

    The ability to capture an acoustic scene in a spatially selective manner is a challenging and centric task in acoustic signal processing. 
    This task is found in many applications, such as spatial sound capturing and reproduction \cite{7038281, thiergart2017robust}, speaker extraction \cite{beamformer_guided_tse1, ftjnf}, and automatic spatial gain control \cite{agc}. 
    
    Beamformers achieve spatial selectivity by linearly filtering multiple microphone signals \cite{Bensty09}. These filters can be obtained, e.g., as a solution to a constrained optimization problem.
    Other works explored hybrid neural beamforming approaches where a conventional beamformer is supported by a \ac{DNN}, e.g., \cite{ADLMVDR}.
    To influence the resulting spatial selectivity, \acp{DNN} have been used to estimate filter vectors \cite{beampattern1, neural_beamforming4, dsenet}, some allowing for control over beamwidth and sharpness during training \cite{dsenet}.

    By relying on parametric sound field models, parametric directional filtering methods, e.g., \cite{pulkki2007spatial, Kallinger09, thiergart-baseline, 7038281}, estimate the \acp{DOA} of active sources and subsequently apply corresponding gains or spatial filters to achieve a desired spatial selectivity pattern.
    However, these methods are often hindered by challenging acoustic conditions that violate their sound field models \cite{parametric_limitation}.

    Alternatively, multi-channel source separation and extraction methods approach spatial selectivity by adopting source-dependent selectivity patterns.
    Particularly interesting are methods that allow for side information to exert limited control over the resulting selectivity pattern.
    Some studies investigated the inclusion of the \ac{DOA} of a target as supplementary information for \acp{DNN}, either for spatial source extraction \cite{neural_beamforming, location_guided1, neural_beamforming5} or for \ac{DNN}-based source separation with a \ac{DOA}- and distance-based loss \cite{location_guided4, location_guided5}.
    Other methods have explored defining spatial selectivity patterns as regions around the array wherein active sources are extracted.
    This includes angular regions with respect to the array \cite{rezero, location_guided2, deep_zoom, location_guided3, location_guided6}, circular/spherical regions around the array \cite{rezero, location_guided3, location_guided8}, a combination of the two \cite{rezero}, or user-defined angular ranges from which to extract sources during inference \cite{deep_zoom}.
    Notable among these methods is the FT-JNF \cite{ftjnf,ftjnf_steerable, tesch_insights, ftjnf_journal}, which utilizes \ac{DNN}-estimated spatial-spectral filters to extract a single target speaker based on an input \ac{DOA}.
   
    Finally, other methods have been proposed to improve spatial selectivity using virtual microphones. They aim to estimate the signal of a virtual microphone at a spatial position where no physical microphone captures the sound field during inference, and then use the physical and virtual signals to achieve better spatial selectivity. These methods have progressed from interpolation-based methods to DNN-based methods \cite{vme1, vme2, vme3}. Nonetheless, none of these methods investigate virtual microphones with different selectivity patterns. 

    However, existing methods for spatial selectivity often lack fine-grained control over the resulting pattern and/or adopt a source-dependent definition of the desired pattern. This renders applications such as spatial sound capturing and reproduction especially challenging, as deviations from the desired pattern can result in a distorted spatial image. 
    While some techniques allow for adjustment of the beam width \cite{dsenet, deep_zoom}, it remains unclear whether a detailed control over the pattern can be achieved.
	
    To investigate this question, this paper presents a preliminary study on the feasibility of employing \acp{DNN} for selectivity pattern learning, in particular focusing on time-invariant patterns. The contributions of this paper are: (i) formalizing the problem of learning a selectivity pattern; (ii) investigating the appropriate composition of training datasets; (iii) demonstrating that \ac{DNN}-based pattern learning enables the realization of selectivity patterns with fewer microphones than classical signal processing.

	\section{Problem Formulation}
	\label{sec:problem_formulation} 

    Consider a scenario in which a small $Q$-microphone array consisting of omnidirectional microphones captures an anechoic acoustic scene with $N$ sound sources located in the far field of the array.
   Let $X_{q,n}[f,t]$ be the $n$-th source signal at the $q$-th microphone in the \ac{STFT} domain, where $f$ and $t$ denote the frequency and time indices, respectively. The $q$-th microphone signal can be written as
    \vspace*{-0.15cm}
    \begin{equation} \label{eqn:mic_sig}
        Y_q[f,t] = \sum_{n=1}^{N} X_{q,n}[f,t] + V_q[f,t],~q\in\{1,2,\ldots,Q\},
    \end{equation}
    where the sensor noise $V_q[f,t]$ is uncorrelated across the microphones and $X_{q,n}[f,t]= H_{\mathbf{p}_q,n}[f] \, X_{n}[f,t]$, where $H_{\mathbf{p}_q,n}[f]$ models the direct-path transfer function (DPTF) between the $n$-th source $X_{n}[f,t]$ and the $q$-th microphone at position $\mathbf{p}_q$.

    We consider the task of {\it directional filtering}, where the goal is to capture the $N$ sources at a position $\mathbf{p}_\textrm{VDM}$ with a specific directivity pattern $S[\vartheta_n,f ]$, where $\vartheta_n$ denotes the \ac{DOA} of the $n$-th source with respect to $\mathbf{p}_\textrm{VDM}$. We refer to the target signal as the \ac{VDM} signal $Z_{\textrm{VDM}}[f,t]$ expressed as
    \vspace*{-0.15cm}
    \begin{equation}\label{eqn:vdm_sig}
        Z_{\textrm{VDM}}[f,t] = \sum_{n=1}^{N} S[\vartheta_n, f] \, H_{{\mathbf{p}_{\textrm{VDM}}},n}[f] \, X_n[f,t].
    \end{equation}
    In the following, $S[\vartheta,f]$ is only a function of the azimuth $\vartheta$ as we restrict the formulation to a two-dimensional pattern learning for the sake of simplicity.

    Parametric filtering approaches assuming a single-plane wave per time-frequency (c.f. \cite{7038281}) estimate $Z_{\textrm{VDM}}[f,t]$ by first assigning each \ac{STFT} bin with a \ac{DOA} $\vartheta[f, t]$ and then scaling a reference microphone signal (here chosen to be the first microphone) with a direction-dependent real-valued gain $G[f, t]$, i.e.,
    \begin{equation}
        \widehat{Z}_{\textrm{VDM}}[f,t] = G[f, t] \, Y_{1}[f,t],
        \label{eq:gain_application}
    \end{equation}
    where $G[f, t]$ is obtained by evaluating the desired directivity pattern $S[\vartheta, f]$ at the estimated DOA $\vartheta[f, t]$. The effectiveness of this method hinges on fulfilling the W-disjoint orthogonality assumption \cite{w_disjoint_orthogonality, parametric_limitation}, which is often violated for scenes containing multiple sources. In addition, the effectiveness depends on the accuracy of the DOA estimates used to compute the direction-dependent gain. In this work, we propose a data-driven approach wherein $Z_{\textrm{VDM}}[f,t]$ is estimated using a \ac{DNN}.

\section{Proposed Method}
\label{sec:proposed_method}
    \subsection{Architecture}\vspace{-0.5em}
    \label{ssec:architecture}
    We adopt the FT-JNF architecture, proposed in \cite{tesch_insights} to extract a single speaker from a particular direction, due to its similarity with our directional filtering task.
    In this architecture, the real and imaginary parts of the $Q$ microphones are stacked along the channel dimension and first fed to a bidirectional \ac{LSTM} layer configured to model the spectro-spatial relationships in the input, whose output is then fed to a second LSTM layer configured to model the temporal relations.
    In this work, we focus on a frame-level causal scenario; therefore, the second LSTM is chosen to be unidirectional.
    Finally, the output of the second LSTM is fed to a linear layer with a hyperbolic Tangent activation function, which computes a complex-valued single-channel mask $\mathcal{M}[f,t]$.
    The desired \ac{VDM} signal is then estimated by applying the mask to the reference microphone signal,
    \begin{equation}\label{eqn:dnn_op}
        \widehat{Z}_{\textrm{VDM}}[f,t] = \mathcal{M}[f,t] Y_{1}[f,t].
    \end{equation}

    \subsection{Loss function}\vspace{-0.5em}
    \label{ssec:loss_function}
    Let $\widehat{z}_{\textrm{VDM}}$ and ${z}_{\textrm{VDM}}$ be the time-domain signals corresponding to the \ac{STFT} representations $\widehat{Z}_{\textrm{VDM}}[f,t]$ and ${Z}_{\textrm{VDM}}[f,t]$, respectively. 
    We consider a distortion measure computed between $\widehat{z}_{\textrm{VDM}}$ and ${z}_{\textrm{VDM}}$ as the loss function.
    The mask $\mathcal{M}[f,t]$ depends on the \acp{DOA} of speakers, and it can have a high dynamic range.
    For example, the reference microphone signal $Y_1[f,t]$ needs to be attenuated significantly if all the speakers in the scene are positioned near the nulls in the directivity pattern.
    Such large attenuation requirements typically lead to a higher dynamic range for the computed gradients and cause instabilities during training.
    Hence, we choose the \ac{tsdr} \cite{sasdr} as the loss function, as it is well-defined for both perfect reconstruction and silence.

    \subsection{Training strategy and simulations}\vspace{-0.5em}
    \label{ssec:simulation_of_mic_signals}
    Our preliminary studies have shown that the learned directivity pattern depends on the number and placement of the speakers in the training dataset.
    We hypothesize that the training dataset must include multi-speaker acoustic scenes with speaker positions densely sampling the desired directivity pattern.
    Since the goal is to learn a far-field directivity pattern in free-field, we restrict the set of admissible speaker positions to a circle with a sufficiently large diameter $d_{\textrm{activity}}$ concentric and coplanar with the microphone array.
    
    To simulate an acoustic scene with $N$ speakers, we (i) randomly select $N$ \acp{DOA}, (ii) simulate the DPTF $H_{\mathbf{p}_q,n}$ for all $N$ speakers and $Q$ microphones using \cite{RIRGenerator} with reflection order zero, and (iii) compute the receive microphone signals using (\ref{eqn:mic_sig}).
    To generate the \ac{VDM} signal, the source signals at the position of the \ac{VDM} are further scaled using the desired direction-dependent gains prior to the summation, as in \eqref{eqn:vdm_sig}.

	\begin{figure}
		\centering
		\resizebox{0.28\textwidth}{!}{
		\begin{tikzpicture}
			\begin{polaraxis}[
				xticklabel=$\pgfmathprintnumber{\tick}^\circ$,
				xtick={0,30,...,330},
				ytick={-50, -40,...,0},
				ymin=-60, ymax=0,
				y coord trafo/.code=\pgfmathparse{#1+60},
				y coord inv trafo/.code=\pgfmathparse{#1-60},
				yticklabel style={anchor=south east, yshift=-0.3cm, xshift=-0.42cm, font=\normalsize, rotate=-90},
				xticklabel style={font=\large},
				y axis line style={opacity=0, yshift=0cm},
				ytick style={opacity=0, yshift=0cm},
				legend to name=fred
				]
				\addplot [no markers, thick, blue] table [col sep=comma] {figures/cardioid_discretized.csv};
				\addlegendentry{Cardioid}
				\addplot [no markers, thick, alorange] table [col sep=comma] {figures/cdma_discretized.csv};
				\addlegendentry{$3^{\textrm{rd}}$-order \acs{DMA}}
			\end{polaraxis}
			\coordinate (c3) at (8,6);
			\node[above] at (c3)
			{\pgfplotslegendfromname{fred}};
		\end{tikzpicture}
		}
        \vspace*{-0.36cm}
		\caption{Logarithmic polar plots of the \acs{DMA} patterns considered in this study.}
		\label{fig:dma_patterns}
        \vspace*{-0.6cm}
	\end{figure}
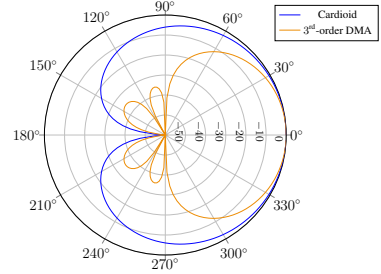
	
	\section{Experimental Setup}
	\label{sec:experimental_setup}

    \subsection{Virtual microphone directivity patterns}\vspace{-0.5em}
    \label{ssec:patterns}

    In this study, two directivity patterns (or, equivalently, two \acp{VDM}) are investigated that could be implemented as \acp{DMA} in practice.
    Generally, the directivity pattern of a ${R\textrm{-th}}$ order \ac{DMA} steered towards $\vartheta_0$ can be expressed as \cite{Elko2004}
    \vspace*{-0.15cm}
	\begin{equation}
		S\left[\vartheta,f\right] = \sum_{r=0}^{R} a_r \cos^r\!\left(\vartheta-\vartheta_0\right) \quad \forall f.
		\label{eq:dma_function}
	\end{equation}
	The first directivity pattern considered in this study is a first-order cardioid pattern with parameters $a_0=\frac{1}{2}, a_1=\frac{1}{2}$ that could be realized as a \ac{CDMA} using three microphones \cite{cdma_book}.
	The second directivity pattern considered in this study is a third-order pattern with parameters $a_0=0, a_1=\frac{1}{6}, a_2=\frac{1}{2}, a_3=\frac{1}{3}$ that could be realized as a \ac{CDMA} using six microphones \cite{cdma_book}.
	Polar plots of the respective patterns are depicted in Fig.~\ref{fig:dma_patterns}.
	
	Note that \acp{CDMA} can be steered into any direction in the plane $\vartheta\in\left[0,2\pi\right]$.
	For comparability, we use $\vartheta_0 = 0$ for both patterns.
 
	\subsection{Array Geometry and Datasets}\vspace{-0.5em}
	\label{ssec:dataset}
	We used a $Q=4$ microphone setup, with three microphones forming a \ac{UCA} and an additional microphone ($q=1$) at the array center. The diameter of the \ac{UCA} was $3\,\textrm{cm}$, which corresponds to a spatial aliasing frequency of $11.4\,\textrm{kHz}$. The \ac{VDM} was placed at the center microphone position, i.e., ${\bf p}_\textrm{VDM} = {\bf p}_1$.


    We simulated the acoustic scenes using the strategy described in Sec. \ref{ssec:simulation_of_mic_signals}. Apart from being coplanar and concentric with the \ac{UCA} on a circle with diameter ${d_{\textrm{activity}}=3\,\textrm{m}}$, the admissible speaker positions were also restricted to $P=144$~equidistant \acp{DOA}. \Acp{DPIR} between the $P$ source positions and the $Q$ microphone positions were first simulated and stored. 
    These simulated \acp{DPIR} were split into three position-disjoint subsets for training, testing, and validation as follows: $P_\textrm{training}=36$ with $\vartheta\in\left\{0^\circ, 10^\circ, \ldots, 350^\circ\right\}$; $P_\textrm{validation}=36$ with $\vartheta\in\left\{5^\circ, 15^\circ, \ldots, 355^\circ\right\}$; $P_\textrm{testing}=72$ with $\vartheta\in\left\{2.5^\circ, 7.5^\circ, \ldots, 357.5^\circ\right\}$. For the \ac{VDM}, we simulated the two patterns described in Sec.~\ref{ssec:patterns}.
	
	As a basis for our datasets, we used the LibriSpeech database \cite{librispeech}.
	Using the subsets `train-clean-360' and `dev-clean', we created training and validation sets with a maximum number of ${N_{\textrm{train}}\in\left\{1,2,3,4,5,6\right\}}$ concurrently active speakers by convolving segments of $4\,\textrm{seconds}$ with the \acp{DPIR} described earlier at a minimum angular separation of $10\,\textrm{degrees}$.
    The number of speakers per sample is chosen uniformly at random.
    Using the subset `test-clean', we created test sets with a fixed number of ${N_{\textrm{test}}\in\left\{1,2,3,4,5,6\right\}}$ speakers in the same way.
    All signals were normalized to have $\left[-33, -25\right]$ loudness units relative to full scale \cite{loudness} after convolution with the \ac{RIR}, similar to \cite{librimix}.
    Samples from LibriSpeech shorter than $4\,\textrm{seconds}$ were extended through zero-padding, with the padding randomly distributed between the beginning and the end of the signal.
	Each training, validation, and test set consists of $10'000$, $3'000$, and $3'000$ samples, respectively.
	
	For the microphones of the array, we also simulated microphone self-noise by adding spatio-temporal white Gaussian noise at a \acl{SNR} of $30\,\textrm{dB}$ with respect to the mixture of all speakers.
	The target \ac{VDM} signal remains noise-free.
		
	\subsection{Training details}\vspace{-0.5em}
	\label{ssec:model_details}
	
	For both target patterns, we trained a total of $6$~models of the FT-JNF architecture \cite{ftjnf} each with an increasing maximum number of speakers ${N_\textrm{train}}$.
	All models were trained for $250$~epochs with a batch size of $10$ and a learning rate of $0.001$. The final model was selected as the model instance whose validation loss (negative \acs{SDR}) was the lowest over all epochs.
	As the number of parameters of the \ac{DNN} only depends on the number of input channels but not on the number of speakers, all trained \acp{DNN} have $873$K~parameters.
	The \ac{STFT} was calculated on signal frames of $32\,\textrm{ms}$ using a $\sqrt{\textrm{Hann}}$ window and $50\,\%$ overlap at a sampling frequency of $16\,\textrm{kHz}$. The threshold and $\varepsilon$ values for the \ac{tsdr} loss function were chosen to be $30$~dB and $1.2\cdot 10^{-7}$, respectively.

    \begin{table}
        \setlength\extrarowheight{0.1pt}
		\caption{\ac{SDR} values [dB] of the reference microphone, baselines, and our proposed method based on FT-JNF \cite{tesch_insights} for the cardioid experiment. The best results per column are highlighted in boldface. The column ``av.'' gives the average per row.}
		\resizebox{.485\textwidth}{!}{
			\begin{tabular}{l c rrr rrr r}
				\toprule
				\multicolumn{1}{c}{\multirow{2}{*}{\textbf{cardioid}}}& & \multicolumn{7}{c}{Number of speakers during testing}\\
				 &\diagbox[innerwidth=0.85cm]{$N_\textrm{train}$}{$N_\textrm{test}$} &\multicolumn{1}{c}{1}&\multicolumn{1}{c}{2}&\multicolumn{1}{c}{3}&\multicolumn{1}{c}{4}&\multicolumn{1}{c}{5}&\multicolumn{1}{c}{6}&av. \\
 				\midrule
				 Reference Microphone & &{\footnotesize -1.8} &-{\footnotesize 0.3} &{\footnotesize -0.2} &{\footnotesize 0.0} &{\footnotesize 0.1} &{\footnotesize 0.0} &{\footnotesize -0.4}\\
				\midrule
                LS Beamformer \cite{ls_beamforming} & &6.3 &10.9 &12.0 &12.5 &12.7 &12.8 &11.2 \\
				Parametric Baseline \cite{7038281} & &27.7 &18.7 &15.5 &14.0 &12.9 &12.0 &16.8 \\
				\midrule
				\multirow{6}{*}{\makecell[l]{FT-JNF \cite{tesch_insights}\\\textit{given the max.}\\\textit{number of speakers}\\\textit{used for training}}}&1 &\textbf{32.2} &13.5 &10.0 &8.8 &8.1 &7.6 &13.4 \\
				&2 &\textbf{32.2} &\textbf{27.9} &25.1 &23.8 &23.0 &22.3 &25.7 \\
				&3 &30.9 &27.8 &26.0 &25.0 &24.2 &23.5 &\textbf{26.2} \\
				&4 &30.3 &27.8 &26.1 &25.1 &24.3 &23.6 &\textbf{26.2} \\
				&5 &30.2 &27.8 &\textbf{26.2} &\textbf{25.2} &\textbf{24.4} &\textbf{23.8} &\textbf{26.2} \\
				&6 &29.2 &27.3 &25.8 &25.0 &24.2 &23.6 &25.8 \\
				\bottomrule
			\end{tabular}
		}
		\label{tab:sdri_results_cardioid}
	\end{table}
	
	\begin{table}
        \setlength\extrarowheight{0.25pt}
		\caption{\ac{SDR} values [dB] of the reference microphone, baselines, and our proposed method based on FT-JNF \cite{tesch_insights} for the $3^{\textrm{rd}}$-order \ac{DMA} experiment. The best results per column are highlighted in boldface. The column ``av.'' gives the average per row.}
		\resizebox{.485\textwidth}{!}{
			\begin{tabular}{l c rrr rrr r}
				\toprule
				\multicolumn{1}{c}{\multirow{2}{*}{\textbf{$3^{\textrm{rd}}$-order \ac{DMA}}}}& &\multicolumn{7}{c}{Number of speakers during testing}\\
				&\diagbox[innerwidth=0.85cm]{$N_\textrm{train}$}{$N_\textrm{test}$} &\multicolumn{1}{c}{1}&\multicolumn{1}{c}{2}&\multicolumn{1}{c}{3}&\multicolumn{1}{c}{4}&\multicolumn{1}{c}{5}&\multicolumn{1}{c}{6}&av. \\
			    \midrule
				Reference Microphone & &{\footnotesize -21.5} &{\footnotesize -15.8} &{\footnotesize -12.0} &{\footnotesize -9.7} &{\footnotesize -8.3} &{\footnotesize -7.6} &{\footnotesize -12.5}\\
				\midrule
                LS Beamformer \cite{ls_beamforming} & &-16.4 &-9.0 &-5.0 &-2.7 &-1.4 &-0.7 &-5.9 \\
				Parametric Baseline \cite{7038281} & &\textbf{25.6} &14.1 &10.9 &9.5 &8.5 &7.8 &12.7 \\
				\midrule
				\multirow{6}{*}{\makecell[l]{FT-JNF \cite{tesch_insights}\\\textit{given the max.}\\\textit{number of speakers}\\\textit{used for training}}}
				&1 &17.3 &7.9 &5.5 &4.6 &4.0 &3.5 &7.1 \\
				&2 &21.1 &\textbf{20.1} &16.3 &14.0 &12.4 &11.4 &15.9 \\
				&3 &18.3 &19.2 &18.9 &17.8 &16.5 &15.2 &17.7 \\
				&4 &18.8 &19.4 &\textbf{19.3} &18.6 &17.6 &16.5 &18.3 \\
				&5 &18.6 &19.5 &\textbf{19.3} &18.6 &17.7 &16.7 &\textbf{18.4} \\
				&6 &17.4 &19.2 &\textbf{19.3} &\textbf{18.8} &\textbf{17.9} &\textbf{17.0} &18.2 \\
				\bottomrule
			\end{tabular}
		}
		\label{tab:sdri_results_cdma}
	\end{table}
 
	\section{Performance Evaluation}
	\label{sec:performance_evaluation}
	
	\subsection{Performance metrics}\vspace{-0.5em}
	\label{ssec:performance_metrics}
    As conventional directivity patterns cannot be obtained for single-channel, signal-dependent masking approaches (such as the one proposed in this work), we rely on a proxy metric to quantify how well the desired pattern is approximated.
    To this end, we utilize a signal error-based measure.
    Specifically, we use the \ac{SDR} between $Z_{\textrm{VDM}}$ and $\widehat{Z}_{\textrm{VDM}}$ to \textit{quantify the distance} between the desired signal/pattern and the obtained one.
    This includes calculating the \ac{SDR} of the estimates as well as the \acl{noisySDR}. 

	Furthermore, we visualize the output pattern as realized by the \ac{DNN}.
	To plot the learned patterns, the estimated mask $\mathcal{M}$ is applied \textit{separately} to the individual direct sounds of the source signals at the reference microphone $q=1$, $X_{1, n}$, and each masked signal is used to estimate the attenuation in the corresponding source's direction.
	
	Note, however, that the mask is not devised to extract single sources from the mixture.
	Further, this separate application of the mask is only possible for simulated data as the direct sound is unobservable in practice.
	Similarly to what is done in conventional spatial filtering (see, e.g., \cite{microphone_arrays}), the estimate of the spatial pattern is realized as the square root of the power ratio before and after masking, i.e.,
    \vspace*{-0.3cm}
	\begin{equation}
		\widehat{S}\left[ \vartheta_n\right] = \sum_{f=1}^{F} \sqrt{\frac{\frac{1}{T}\sum_{t=1}^{T}\left| \mathcal{M}[f, t] \; X_{1, n}[f, t] \right|^2}{\frac{1}{T}\sum_{t=1}^{T}\left| X_{1, n}[f, t] \right|^2}}.
		\label{eq:pattern_estimate}
	\end{equation}
	
	\subsection{Performance analysis}\vspace{-0.5em}
	\label{ssec:performance_analysis}
	
	Hereinafter, we analyze the performance of the proposed method in detail.
	We evaluate two baselines.
	First, we evaluate the performance of a parametric baseline \cite{7038281}.
	For this baseline, single oracle \acp{DOA} per \ac{STFT} bin are computed as a weighted average of the \acp{DOA} of active sources, based on their contribution to the power of the respective bin.
	This procedure mimics the behavior of a ``real'' \ac{DOA} estimator.
	Then, gains according to the desired pattern(s) are applied to the \ac{STFT} bins of the center microphone, cf. (\ref{eq:gain_application}).
 
    Secondly, we assessed the performance of a fixed, signal-independent \ac{LS} beamformer \cite{ls_beamforming} that was optimized individually for each of the two desired patterns at a minimum white noise gain of $-15\,\textrm{dB}$.    
    
    We tested all methods on the test set(s) as described in Sec.~\ref{ssec:dataset} with (fixed) ${N_\textrm{test}\in\left\{1,2,3,4,5,6\right\}}$ concurrently active speakers.
	Table~\ref{tab:sdri_results_cardioid}, summarizes the results of the cardioid experiment. \newline 
    While the mean \ac{SDR} of the reference microphone is close to $0\,\textrm{dB}$, meaning that the power of the difference between the cardioid target and the omnidirectional microphone is close to the power of the target signal itself, the \acp{SDR} clearly indicate that all methods can successfully realize the desired spatial selectivity\footnote{\label{audio_ex}Audio examples can be found at \url{https://www.audiolabs-erlangen.de/resources/2024-IWAENC-NDF}.}.
	
	In Table~\ref{tab:sdri_results_cdma}, we summarize the results for the $3^{\textrm{rd}}$-order \ac{DMA} experiment.
	While the \ac{SDR} values of the reference microphone are generally lower, the findings of the cardioid experiment can be transferred to this experiment.

    While the \ac{LS} beamformer approximates the Cardioid pattern accurately, and the $3^{\textrm{rd}}$-order \ac{DMA} pattern coarsely, its performance is mostly limited by the strong influence of the white noise.%
    While, for the cardioid target, positive \ac{SDR} values are achieved, it falls short of all other tested methods in the mean.
    Further, the \ac{LS} beamformer cannot approximate the $3^{\textrm{rd}}$-order \ac{DMA} pattern well, reflected in negative \ac{SDR} values of the estimate.
    
    The parametric baseline has a very strong performance when testing with one active speaker only.
    For the $3^{\textrm{rd}}$-order \ac{DMA} experiment, it even outperforms all other methods for this test case, achieving an \ac{SDR} of $25.6\,\textrm{dB}$.
    However, the performance drops significantly when testing with two or more speakers.
    While this could be improved by considering the multi-wave model in \cite{thiergart-baseline}, the performance remains limited by the accuracy of the estimated number of active sound sources and their \acp{DOA}.
    
    Training FT-JNF \cite{tesch_insights} with the proposed method works successfully and widely outperforms both baselines with the exception noted earlier and when comparing models trained on a single speaker against the baselines.
    This means, training on the dataset with a single active speaker is not sufficient to fully learn the directivity.
    While FT-JNF trained on one speaker shows very good performance on testing with one speaker (\acp{SDR} of $32.2\,\textrm{dB}$ and $17.3\,\textrm{dB}$ for cardioid and $3^{\textrm{rd}}$-order \ac{DMA}, respectively), the \ac{DNN} does not generalize to the case of two or more concurrently active speakers, the \ac{SDR} performance drops significantly.
    
    In the scope of our experiments, we find that \acp{DNN} trained on two or more concurrently active speakers generalize to scenarios with up to six speakers.
    Importantly, training with more than three speakers does not further improve the results significantly.
    Based on the mean performance of testing with one to six speakers, we chose FT-JNF trained on mixtures of at most five speakers as our best model, exhibiting a mean \ac{SDR} of $26.2\,\textrm{dB}$ for the cardioid experiment and a mean \ac{SDR} of $18.4\,\textrm{dB}$ for the $3^{\textrm{rd}}$-order \ac{DMA} experiment.
    
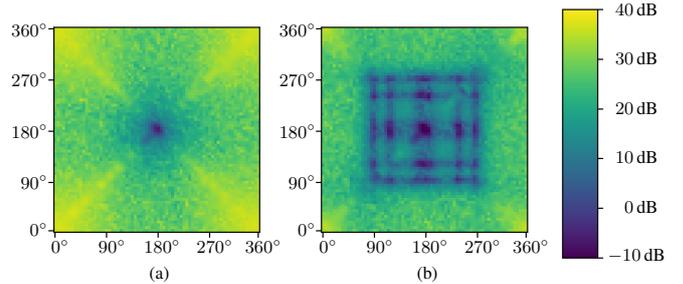
\begin{figure}[!t]
		\centering
		\scalebox{0.75}{
			\begin{tikzpicture}
				\node (sdrmap) at (0, 0) {\includegraphics[width=10cm,height=4.76cm]{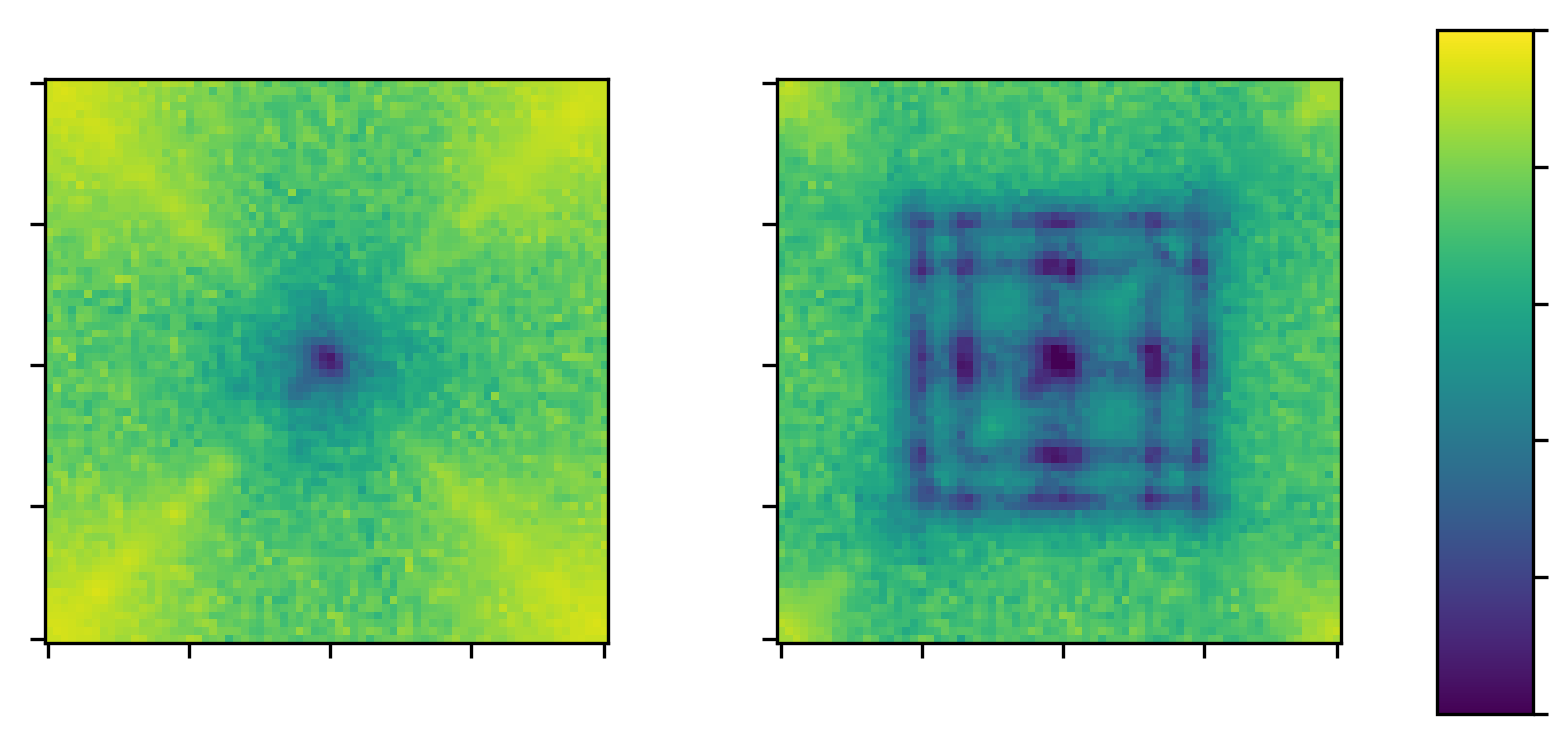}};
				
				\node[] at (-5.15, 1.82) {$360^\circ$};
				\node[] at (-5.15, 0.95) {$270^\circ$};
				\node[] at (-5.15, 0.03) {$180^\circ$};
				\node[] at (-5.15, -0.87) {$\phantom{1}90^\circ$};
				\node[] at (-5.15, -1.7) {$\phantom{18}0^\circ$};
				
				\node[] at (-4.61, -2.0) {$0^\circ$};
				\node[] at (-3.7, -2.0) {$90^\circ$};
				\node[] at (-2.82, -2.0) {$180^\circ$};
				\node[] at (-1.9, -2.0) {$270^\circ$};
				\node[] at (-1.05, -2.0) {$360^\circ$};
				
				\node[] at (-0.45, 1.82) {$360^\circ$};
				\node[] at (-0.45, 0.95) {$270^\circ$};
				\node[] at (-0.45, 0.03) {$180^\circ$};
				\node[] at (-0.45, -0.87) {$\phantom{1}90^\circ$};
				\node[] at (-0.45, -1.7) {$\phantom{18}0^\circ$};
				
				\node[] at (-4.61+4.7, -2.0) {$0^\circ$};
				\node[] at (-3.7+4.7, -2.0) {$90^\circ$};
				\node[] at (-2.82+4.7, -2.0) {$180^\circ$};
				\node[] at (-1.9+4.7, -2.0) {$270^\circ$};
				\node[] at (-1.05+4.7, -2.0) {$360^\circ$};
				
				\node[] at (5.5, 2.18) {$40\,\textrm{dB}$};
				\node[] at (5.5, 1.33) {$30\,\textrm{dB}$};
				\node[] at (5.5, 0.45) {$20\,\textrm{dB}$};
				\node[] at (5.5, -0.42) {$10\,\textrm{dB}$};
				\node[] at (5.5, -1.31) {$\phantom{1}0\,\textrm{dB}$};
				\node[] at (5.5, -2.17) {$-10\,\textrm{dB}$};
				
				\node[] at (-2.87, -2.5) {(a)};
				\node[] at (-2.87+4.7, -2.5) {(b)};
		\end{tikzpicture}}
		\vspace*{-0.7cm}
		\caption{Distributions of \ac{SDR} values when testing the best model for (a) the cardioid target and (b) the $3^{\textrm{rd}}$-order \ac{DMA} target. The models were tested with two concurrently active sources. Their \acp{DOA} are given on the axes, \acp{SDR} for missing \ac{DOA} combinations were interpolated using cubic interpolation.}
		\label{fig:sdr_dists}
		\vspace*{-0.2cm}
	\end{figure}

    For this best model, we further investigate the distribution of \acp{SDR} over the angular range when testing with two speakers.
    In Fig.~\ref{fig:sdr_dists}, we illustrate these distributions.
    It is found that the best \ac{SDR} performance is achieved when both sources are placed close to the \ac{VDM}'s look direction $\vartheta_0=0^\circ$/$360^\circ$.
    There, the omnidirectional microphone is already a good approximation to the target \acp{VDM}.
    
    In contrast, the SDR is notably diminished when both sources are close to the \ac{VDM}'s spatial null, i.e., $180^\circ$ for the cardioid (cf. middle point of Fig.~\ref{fig:sdr_dists}a) and $90^\circ, 120^\circ, 180^\circ, 240^\circ, 270^\circ$ for the $3^{\textrm{rd}}$-order \ac{DMA} (cf. horizontal and vertical lines towards the center of Fig.~\ref{fig:sdr_dists}b).
    This is because the \ac{DNN} needs to strongly process the signals as received by the omnidirectional microphone to approximate the signal of the \ac{VDM}.
    Note that the low \ac{SDR} values there are not only impacted by the models' performance but also by the generally badly conditioned \ac{SDR} for low-power signals and noise-free targets.
    We encourage the reader to listen to the audio examples on the website\textsuperscript{\ref{audio_ex}}, as, perceptually, the signals are well suppressed.
    
    We reason that the spatial one and the spatial null(s) are the edge cases of the task.
    Other combinations of \acp{DOA} exhibit a performance that is found to be rather constant.
    
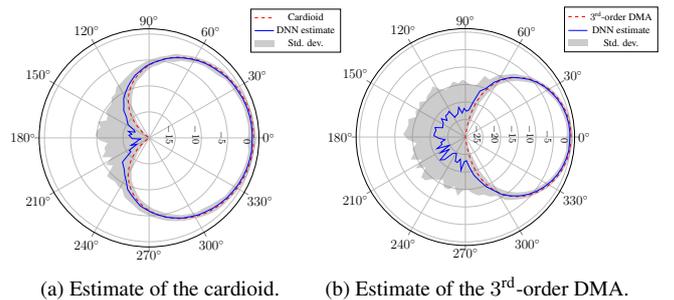
\begin{figure}[!t]
		\centering
        \begin{subfigure}{0.23\textwidth}
            \resizebox{1.1\textwidth}{!}{
            \begin{tikzpicture}
                \begin{polaraxis}[
                    xticklabel=$\pgfmathprintnumber{\tick}^\circ$,
                    xtick={0,30,...,330},
                    ytick={-15, -10,...,0},
                    ymin=-20.0001, ymax=1,
                    y coord trafo/.code=\pgfmathparse{#1+20},
                    y coord inv trafo/.code=\pgfmathparse{#1-20},
                    yticklabel style={anchor=south east, yshift=-0.3cm, xshift=-0.42cm, font=\normalsize, rotate=-90},
                    xticklabel style={font=\large},
                    y axis line style={opacity=0, yshift=0cm},
                    ytick style={opacity=0, yshift=0cm},
                    legend to name=fred2
                    ]
                    \addplot [no markers, thick, red, dashed] table [col sep=comma] {figures/cardioid_discretized_capped.csv};
                    \addlegendentry{Cardioid};
                    
                    \addplot [name path=cardioid, no markers, thick, blue] table [col sep=comma] {figures/cardioid_dnn_estimate.csv};
                    \addlegendentry{DNN estimate};

                    
                    \addplot [name path=cms, no markers, black, opacity=0] table [col sep=comma] {figures/cardioid_dnn_estimate_minus_std.csv};
                    \addplot [name path=cps, no markers, red, opacity=0] table [col sep=comma] {figures/cardioid_dnn_estimate_plus_std.csv};
                    \addplot[gray, opacity=0.4] fill between[of=cms and cps];
                    \addlegendentry{Std. dev.};
                \end{polaraxis}
                \coordinate (c3) at (8,6);
                \node[above] at (c3)
                {\pgfplotslegendfromname{fred2}};
                
                \coordinate (c4) at ($(c3) + (-1.28,0.3)$);
                \draw[rectangle, fill=gray, opacity=0.4, draw=none](c4) rectangle ++(0.6cm, 0.2cm) ;
            \end{tikzpicture}
            }
            \caption{Estimate of the cardioid.}
            \label{fig:sub1}
        \end{subfigure}
        \begin{subfigure}{0.23\textwidth}
            \resizebox{1.1\textwidth}{!}{
            \begin{tikzpicture}
                \begin{polaraxis}[
                    xticklabel=$\pgfmathprintnumber{\tick}^\circ$,
                    xtick={0,30,...,330},
                    ytick={-25, -20,...,0},
                    ymin=-30.0001, ymax=1,
                    y coord trafo/.code=\pgfmathparse{#1+30},
                    y coord inv trafo/.code=\pgfmathparse{#1-30},
                    yticklabel style={anchor=south east, yshift=-0.3cm, xshift=-0.42cm, font=\normalsize, rotate=-90},
                    xticklabel style={font=\large},
                    y axis line style={opacity=0, yshift=0cm},
                    ytick style={opacity=0, yshift=0cm},
                    legend to name=fred3
                    ]
                    \addplot [no markers, thick, red, dashed] table [col sep=comma] {figures/cdma_discretized_capped.csv};
                    \addlegendentry{$3^\textrm{rd}$-order DMA};
                    
                    \addplot [name path=cardioid, no markers, thick, blue] table [col sep=comma] {figures/cdma_dnn_estimate.csv};
                    \addlegendentry{DNN estimate};
                    
                    \addplot [name path=cms, no markers, black, opacity=0] table [col sep=comma] {figures/cdma_dnn_estimate_minus_std.csv};
                    \addplot [name path=cps, no markers, red, opacity=0] table [col sep=comma] {figures/cdma_dnn_estimate_plus_std.csv};
                    \addplot[gray, opacity=0.4] fill between[of=cms and cps];
                    \addlegendentry{Std. dev.};
                \end{polaraxis}
                \coordinate (c3) at (8,6);
                \node[above] at (c3)
                {\pgfplotslegendfromname{fred3}};
                
                \coordinate (c4) at ($(c3) + (-1.34,0.3)$);
                \draw[rectangle, fill=gray, opacity=0.4, draw=none](c4) rectangle ++(0.6cm, 0.2cm) ;
            \end{tikzpicture}
            }
            \caption{Estimate of the $3^{\textrm{rd}}$-order \ac{DMA}.}
            \label{fig:sub2}
        \end{subfigure}
		\caption{Polar plot of the two \ac{VDM} targets and their corresponding \ac{DNN} estimate. The gray area illustrates the standard deviation.}
		\label{fig:polar_plots_dnn}
        \vspace*{-0.6cm}
	\end{figure}

    Finally, again for the best models and testing with two concurrent sources, we compute the patterns as realized by the \ac{DNN} according to (\ref{eq:pattern_estimate}) and illustrate the results in Figs.~\ref{fig:sub1}~and~\ref{fig:sub2}.
    The plots are given in logarithmic representation alongside the \ac{VDM}'s target patterns.
    Note that the radial axes are scaled differently between the plots in this paper to match the dynamic range of the depicted patterns.
    For the cardioid, the \ac{DNN} estimate follows the target pattern with minimal deviations and small standard deviation for gain values of up to $-7.5\,\textrm{dB}$, for the $3^{\textrm{rd}}$-order \ac{DMA} up to $-15\,\textrm{dB}$.
    For higher attenuation values, the deviations increase.
    Still, the spatial null is within one standard deviation from the mean.
    The left semi-circle of the $3^{\textrm{rd}}$-order \ac{DMA} estimate is attenuated with around $-22.5\,\textrm{dB}$.
    The lobes as seen in Fig.~\ref{fig:dma_patterns} are below this level.
	
    \vspace{-1em}
	\section{Conclusion}
	\label{sec:conclusion}
    \vspace{-0.5em}

    We employed a \ac{DNN}-based method to process signals from a small array of omnidirectional microphones, approximating the pattern of a \ac{VDM} through implicit learning from data.
    This approach utilizes a recently proposed small, causal \ac{DNN}.
    Our findings demonstrate that this method surpasses traditional parametric and beamforming strategies.
    Notably, we observe significant enhancements in signal quality for both \ac{VDM} training targets.
    Further, we provided insights into the impact of training dataset composition on the performance of the \ac{DNN}.
    Our study highlights the potential of neural directional filtering for far-field directivity control.
    Future work includes exploring steerable and arbitrary patterns, assessing the performance in near-field and reverberant scenarios, testing on measured data, extending the method to elevated sources, and investigating the placement of the \ac{VDM}.

    \balance
	\bibliographystyle{IEEEbib}
	\bibliography{refs}
	
\end{document}